\documentclass{iaus}
\usepackage{graphicx}

\title[Star-forming regions in the intragroup medium of compact groups of galaxies] %% give here short title %%
{Star-forming regions in the intragroup medium of compact groups of galaxies}

\author[S. Torres-Flores et al.]   %% give here short author list %%
{S. Torres-Flores$^{1,2}$, C. Mendes de Oliveira$^{1}$, D. F. de Mello$^{3,4}$, P. Amram$^{2}$, H. Plana$^{2,6}$, B. Epinat$^{5}$ \and J. Iglesias-P\'aramo$^{7}$}
\affiliation{$^1$Universidade de S\~ao Paulo, Instituto de Astronomia, Geof\' isica e Ci\^encias Atmosf\'ericas, Departamento de Astronomia, S\~ao Paulo, Brazil \\ email: {\tt storres@astro.iag.usp.br} \\[\affilskip]
$^2$Laboratoire d'Astrophysique de Marseille, OAMP, Universit\'e de Provence \& CNRS, 38 rue F. Joliot--Curie, 13388 Marseille, Cedex 13, France \\[\affilskip]
$^3$Observational Cosmology Laboratory, Code 665, Goddard Space Flight Center, Greenbelt, MD 20771, USA \\[\affilskip]
$^4$Catholic University of America, Washington, DC 20064, USA \\[\affilskip]
$^5$Laboratoire d'Astrophysique de Toulouse-Tarbes, Universit\'e de Toulouse, CNRS, 14 Avenue Edouard Belin, 31400 Toulouse, France \\[\affilskip]
$^6$Laboratorio de Astrofisica Teorica e Observacional, Universidade Estadual de Santa Cruz, Brazil \\[\affilskip]
$^7$Instituto de Astrofisica de Andalucia (CSIC), Camino Bajo de Huetor 50, 18008 Granada, Spain}

\pubyear{2009}
\volume{262}  %% insert here IAU Symposium No.
\pagerange{1--2}
\jname{Stellar Populations: Planning for the Next Decade}
\editors{G. Bruzual et al.}
\begin{document}

\maketitle

\begin{abstract}

We present the results of a multiwavelength campaign searching for young objects in
the intragroup medium of seven compact groups of galaxies: HCG 2, 7, 22, 23, 92, 100
and NGC 92. We used Fabry-Perot velocity fields and rotation curves together with
GALEX NUV and FUV images, optical R-band and HI maps to evaluate the stage of
interaction of each group. We conclude that groups (i) HCG 7 and HCG 23 are in
an early stage of interaction, (ii) HCG 2 and HCG 22 are mildly interacting, and (iii)
HCG 92, HCG 100 and NGC 92 are in a late stage of evolution. Evolved groups have
a population of young objects in their intragroup medium while no such population
is found within the less evolved groups. We also report the discovery of a tidal dwarf
galaxy candidate in the tail of NGC 92. These three groups, besides containing galaxies
which have peculiar velocity fields, also show extended HI tails. Our results indicate
that the advanced stage of evolution of a group together with the presence of intragroup
HI clouds may lead to star formation in the intragroup medium.

\end{abstract}

\firstsection % if your document starts with a section,
\section{Introduction}

Compact groups of galaxies are associations of three to seven galaxies, where the projected distances between them is of the order of their diameters, and where the group
shows a low velocity dispersion, making compact groups an ideal place to study
galaxy interaction and intergalactic star formation (e.g. \cite[Torres-Flores et
al. 2010]{tor09b}, \cite[de Mello et al. 2008]{dem08b}, \cite[de Mello, Torres--Flores \& Mendes de Oliveira 2008]{dem08a}, \cite[Mendes de Oliveira et al. 2004]{mdeo04}). The main goal of this work is to search for a link between the evolutionary stage of a group and the presence of young intergalactic objects which may have formed during galaxy interactions. For this, we analyze a subsample of seven compact groups (HCG 2, 7, 22, 23, 92, 100 and NGC 92) which span a wide range of evolutionary stages, from HI rich groups to strongly interacting groups, where the galaxies show tidal tail features and a deficiency in neutral HI gas. In order to analyze the evolutionary stage of each group, we used new Fabry-Perot velocity maps, GALEX/UV data and optical R-band images. The velocity fields and rotation curves help constraining the evolutionary stage of each compact group while ultraviolet light contains important information regarding the age of the young stellar population that may be present in the intragroup medium.

\section{UV analysis}

We searched for ultraviolet emitting regions in the vicinity of all seven targets, using the SExtractor software (SE, \cite[Bertin \& Arnouts 1996]{ber96}) in the FUV, NUV and R sky-subtracted images of our compact group sample. We compare the field density of regions detected in the compact group with a control sample outside the group. HCG 92 and HCG 22 have the highest field density in this study. No excess was found in HCG 2, HCG 7, HCG 23, HCG 100 and NGC 92  (\cite[Torres-Flores et al. 2009]{tor09}).

\section{Fabry-Perot analysis}

In order to constrain the evolutionary stage of each compact group, we
inspected the velocity field and rotation curve of each galaxy to search for
interaction indicators, in a similar way to that done by \cite[Plana et al. (2003)]{pla03} and \cite[Amram et al. (2003)]{amr03}. In the case of NGC 92, it shows a prominent tidal tail in its velocity field. At the tip of this tail, there is a tidal dwarf galaxy candidate, having an age of about 40 Myrs (\cite[Torres-Flores et al. 2009]{tor09}).

\section{Conclusions}

We used multiwavelength data to study the evolutionary stages of the compact groups of
galaxies HCG 2, 7, 22, 23, 92, 100 and NGC 92. New Fabry-Perot velocity fields, rotation curves and GALEX NUV/FUV images were analyzed for four and seven of these groups respectively. Groups HCG 7 and 23 are in an early stage of interaction whereas
HCG 2 and 22 show limited interaction features and HCG 92, 100 and NGC 92 are in a
late stage of evolution, having HI gas in the intragroup medium, galaxies with peculiar velocity fields and several young star-forming regions in the intergalactic medium.

\acknowledgments

S. T--F. acknowledges the financial support of FAPESP through the Doctoral position, under contract 2007/07973-3 and Eiffel scholarship.

\end{document}